# The mean-field dividing interface is united with the Widom line

Hongqin Liu (刘洪勤)

Integrated High Performance Computing Branch,

Shared Services Canada, Montreal, Canada

Emails: hongqin.liu@ssc-spc.gc.ca; hqliu2000@gmail.com

## Abstract

We define a mean-field crossover generated by the Maxwell construction as the dividing interface for the vapor-liquid interface area and a highly accurate density-profile equation is thus derived. By using a mean-field equation of sate for the Lennard-Jones fluid incorporated with the density gradient theory, we show that the intrinsic free energy peaks and the isobaric heat capacity exhibits local maxima at the interface. We demonstrate that the mean-field interface is the natural extension of the Widom line into the coexistence region, hence the entire space is coherently divided into liquid-like and gas-like regions in all three (temperature-pressure-volume) planes. Finally, the mean-field theory is found holding all the information for composing the phase diagrams over the entire phase space.

A vapor-liquid interface system is composed of homogeneous bulk vapor and liquid phases and an inhomogeneous interfacial region. In the interfacial region, the discontinuous vapor and liquid densities are bridged together by a position-dependent local density, $\rho(z)$, known as the density profile, where $z$ is the position variable normal to the interface. As the density profile is known, various properties can be expressed as functions of the local density. A closely related concept is the Gibbs dividing interface, which divides the interface area into two regions. The definition of the dividing interface is somewhat arbitrarily [1,2] and conventionally the arithmetic mean of vapor and liquid densities is used to define the diving interface. Namely, the origin ($z = z_0$) of the interface area is defined at the average density, $\rho(0) = (\rho_v + \rho_L)/2$. The classic density profile function based on this dividing interface suffers a major shortcoming: the decay length (or thickness) of the vapor side is the same as that of the liquid side, while physical observation tells that the decay length of vapor boundary is less than that of liquid boundary [3]. In addition, both the mean-field theory and the density function theory based on the classic profile fail to predict the local maxima of the difference between the normal and tangential components of the pressure tensor [4].

The vapor-liquid interfacial region diminishes as the temperature rises and finally disappears at the critical point. A system above the critical point is known as the supercritical fluid. Traditionally, the supercritical fluid is considered to be a uniform phase. In last decades or so, some special features have been discovered in the region. Outstandingly, the Widom line [5], which is defined with the local maxima of isobaric heat capacity, $C_p$, divides a supercritical area into gas-like and liquid-like regions [6-9]. However, the relationship between the behaviors in the supercritical region and those in the interface area is rarely addressed. Currently, the Widom line is considered to be a smooth continuation of the saturated pressure in the pressure-temperature space, while in the pressure-volume (density) and temperature-volume spaces, the extension of the Widom line becomes bifurcated.

In this work, we revisit the theories for the vapor-liquid interface system, namely the density gradient theory [1,2, 10] and the mean-field theory (equation of state) [2,11]. The objective is threefold: (1) defining a new dividing interface and thus proposing a novel density profile model; (2) exploring the behaviors of the vapor-liquid interface system with the density gradient theory and the mean-field theory with the new density profile model; (3) addressing the relationship between the behaviors in the supercritical region and those in the interfacial region.

According to the density gradient theory [1,2], for an interface system with a planar interface, the total free energy density ($\mathcal{F}$) can be expressed as:

$$\mathcal{F} = \mathcal{A}\int_{-\infty}^{\infty}\left[f_0\big(\rho(z)\big) + \frac{1}{2}m\left(\frac{d\rho}{dz}\right)^2\right]dz \qquad (1)$$

where $\mathcal{A}$ t is the interface area, $f_0\big(\rho(z)\big) = a_0(\rho)\rho$, the free energy density of a homogeneous fluid with density $\rho(z)$, and $m$, the influence parameter. In Eq.(1) the expansion of local free energy density, $f\big(\rho(z)\big) = F\big(\rho(z)\big)\rho(z)$, is around the "homogeneous" value, and all odd-order terms vanish since a change of the variable from $z$ to $z' = -z$ should not alter the values of the free energy [2]. On minimizing the total free energy, Eq.(1), one gets:

$$\frac{m}{2}\left(\frac{d\rho}{dz}\right)^2 = \left[a_0\big(\rho(z)\big) - \mu^s\right]\rho(z) + P^s \qquad (2)$$

where, $\mu(\rho) = df_0/d\rho$, $P^s$ and $\mu^s$ are the pressure and the chemical potential at saturated (equilibrium) condition, respectively. Eq.(2) relates the density profile to local thermodynamic properties. The free energy $a_0\big(\rho(z)\big)$ can be evaluated by a mean-field equation of state (EoS) with $\rho(z)$ as the density. Then the surface tension can be calculated by the following: $\gamma = \sqrt{2m}\int_{\rho_v}^{\rho_L}\sqrt{\Delta\Omega}\,d\rho$, where $\rho_v$ and $\rho_L$ are the saturated densities for the bulk vapor and liquid phases, respectively and the grand potential is given by $\Omega = \left[a_0\big(\rho(z)\big) - \mu^s\right]\rho(z)$. The influence parameter can be estimated from surface tension data. The later is related to the difference of the normal and tangential components of the pressure tensor, $p_N(z) - p_T(z)$. Finally, the pressure difference can be computed from the following:

$$\Delta\Omega = \frac{1}{2}\left[p_N(z) - p_T(z)\right] = \left[a_0\big(\rho(z)\big) - \mu^{*s}\right]\rho(z) + P^{*s} \qquad (3)$$

Given $p_N(z) - p_T(z)$ from computer simulations, Eq.(3) can be employed to test the predictions of the gradient theory together with a mean-field EoS and a density profile. For applications of the density gradient theory, Eq.(1) and Eq.(2) etc., the density profile, $\rho(z)$, and it's derivative, $d\rho/dz$, are required. Several analytical expressions have been proposed for $\rho(z)$ [13] and a simple classic model has been widely employed [1,2,4,13]:

$$\rho(z) = \frac{\rho_v + \rho_L}{2} + \frac{\rho_L - \rho_v}{2}\tanh\left(\frac{z - z_0}{D}\right) \qquad (4)$$

where the parameter $D$ is related to the thickness of the interfacial region with a planar interface [13]. The dividing interface in Eq.(4) is the algebraic mean density. In addition, for a density profile model, the following boundary conditions have to be imposed [1-3]: $z - z_0 = -\infty,\ \rho(z) = \rho_v$; $z - z_0 = \infty,\ \rho(z) = \rho_L$; $\left.\frac{d^n\rho(z)}{dz^n}\right|_{z\to\pm\infty} = 0$. Apparently, Eq.(4) satisfies all the boundary conditions due to the tanh(x) function. The equal boundary "thicknesses", related to the parameter $D/2$ in Eq.(4), at vapor and liquid sides apparently contradicts with the physical expectations [3].

For treating of the vapor boundary and the liquid boundary separately, we have to define a new dividing interface. In solving a vapor-liquid equilibrium (VLE) problem, namely calculating the saturated densities and pressure at a given temperature, a mean-field EoS is usually employed, subjected to the Maxwell construction. A vast majority of equations of state provide three solutions: one as the saturated vapor volume, another as the liquid volume, plus an "unphysical" solution. The last one has been considered "unphysical" since it corresponds to a state in the unstable interfacial region and it has always been discarded in all the calculations. In fact this "unphysical" solution, $\rho_M$, is well defined as a companion of the saturated densities. In the context of the density gradient theory, any state and it's properties can be meaningfully defined in the interfacial region. Some state, i.e. the average density, may have no particular physical significance though. Here we define the trajectory of the third solution, $\rho_M$, denoted as the M-line, as the mean-field dividing interface, which can be generated from the Maxwell construction [14,15], namely:

$$z - z_0 = 0,\ \ \rho(z) = \rho_M \qquad (5)$$

Based on this mean-field dividing interface and a strict analysis on the governing differential equations for interfacial region [3], we are able to derive a new density profile equation that satisfies all the boundary conditions and Eq.(5) (see details in Ref [16]). The final expression reads:

$$\rho(z) = \rho_M + \frac{2(\rho_L - \rho_M)(\rho_M - \rho_v)\tanh\left(\frac{z - z_0}{D_v}\right)}{\rho_L - \rho_v + (2\rho_M - \rho_v - \rho_L)\tanh\left(\frac{z - z_0}{D_L}\right)} \quad (6)$$

The parameters, $D_v$, $D_L$ and $z_0$ can be obtained by fitting the equation with computer simulation data. With two parameters, $D_v$ and $D_L$, the new model separates the vapor side of the interface from the liquid side and thus is physically favorable.

The Lennard-Jones (LJ) fluid is one of the most important model fluids employed in various areas of condensed matter physics. The molecular potential of the LJ fluid is given by the following: $u(r) = 4\epsilon\left[\left(\frac{\sigma}{r}\right)^{12} - \left(\frac{\sigma}{r}\right)^{6}\right]$, where $\epsilon$ and $\sigma$ are the energy and diameter parameters, respectively. For the vapor-liquid interface system, many publications have been devoted to the LJ fluids [17,18] and to the truncated and shifted LJ fluid [4,19]. In this work, we make use of the LJ fluid for quantitative analysis.

The data for the density profile and the pressure difference, $p_N(z) - p_T(z)$, are adopted from the recent computer simulations of Mejía et al. [17,20] for the temperature range, $0.65 \le T^* \le 1.1$. The density profile data are used to fit Eq.(6) with parameters as functions of temperature: $D_L = 3.71168T^{*2} - 3.11467T^* + 1.77358$, $D_v = 6.79815T^{*2} - 8.85396T^* + 3.83015$, $z_0 = 44.82538T^{*3} - 90.6502T^{*2} + 69.70747T^* - 17.8015$. For the classic model, Eq.(4), it's parameters are also obtained by fitting the same data [16].

For calculations of various thermodynamic properties here we employ the mean-field LJ EoS proposed recently by Stephan et al. [11] due to it's high accuracy and relative simplicity. In all the calculations, the reduced quantities are used and indicated with the superscript "*": $\rho^* = N\sigma^3/V$, $T^* = T/k_B\epsilon$, $P^* = P\,\sigma^3/k_B\epsilon$, $a^* = F/\epsilon$, etc. In the mean-field EoS [11], the residual Helmholtz free energy is composed of a reference (the hard sphere fluid) term and a perturbation contribution [11]:

$$a^* = T^*(ln\rho^* + \tilde{a}^{res}) = T^*(ln\rho^* + \tilde{a}^{ref} + \tilde{a}^{pert}) \quad (7)$$

where the term $ln\rho^*$ comes from the ideal gas and a temperature-dependent term is ignored without impacting the final results. The expressions for the reference term $\tilde{a}^{ref}$, the perturbation term $\tilde{a}^{pert}$ and related quantities can be found in Refs [11,16] and omitted here. The intrinsic local free energy is calculated by the following:

$$\boldsymbol{f}[\rho(z)] = \frac{f_0(\rho(z))}{\rho(z)} + \frac{1}{2}\frac{m}{\rho(z)}\left(\frac{d\rho}{dz}\right)^2 = a^*(\rho(z)) + f_2(\rho(z)) \quad (8)$$

where $f_2(\rho(z)) = \frac{1}{2}\frac{m}{\rho(z)}\left(\frac{d\rho}{dz}\right)^2$ is the heterogeneous contribution and the density $\rho(z)$ is introduced to convert the free energy density to the reduced value. After Ref.[2], we define an intrinsic property as the sum of a homogeneous contribution and a heterogeneous (density gradient) contribution. From the derivation processes [3,16], we note that Eq.(8) is exact up to the 3rd-roder term while omitting the 4th and higher orders. Therefore the results obtained below are accurate enough for our analysis. In addition to the intrinsic Helmholtz free energy, Eq.(8), we are also interested in heat capacity. For a heterogeneous interface system, the intrinsic local heat capacity at constant pressure can be calculated from an EoS by the following [4,16]:

$$\boldsymbol{C_P}(\rho(z)) = -\left(\tilde{a}_{20}^{id} + \tilde{a}_{20}^{int}\right) + \frac{\left(1 + \tilde{a}_{01}^{int} - \tilde{a}_{11}^{int}\right)^2}{1 + 2\tilde{a}_{01}^{int} + \tilde{a}_{02}^{int}} \quad (9)$$

where $-\left(\tilde{a}_{20}^{id} + \tilde{a}_{20}^{int}\right) = \boldsymbol{C_v}(\rho(z))$, is the heat capacity at constant volume, the derivatives are defined by the following [25]: $\tilde{a}_{nm}^{int} = \tau^n \rho^{*m}\frac{\partial^{n+m}(\tilde{a}^{res} + \tilde{f}_2)}{\partial\tau^n\partial\rho^{*m}}$, where $\tilde{a} = a/T^*$, $\tau = 1/T^*$, $\tilde{a}_{20}^{id} = 1.5$ and $\tilde{f}_2(\rho(z)) = f_2(\rho(z))/T^*$. The homogeneous contribution to the residual free energy is given by $\tilde{a}^{res} = a_0(\rho(z)) - \tilde{a}^{id}$, and the contribution of the gradient term should be included for evaluating the heat capacity: $\tilde{a}^{int} = \tilde{a}^{res} + \tilde{f}_2$. The values of $\tilde{f}_2$ are first calculated from the density profile, Eq.(6), then the density and temperature dependences are correlated with a polynomial function:

$$\tilde{f}_2(\rho(z)) = \sum_{i=0}^{4} c_{4-i}\rho(z)^{4-i} \quad (10)$$

where $c_j = \sum_{k=0}^{3} c_{j,3-k}\tau^{(3-k)}$, more details are provided elsewhere[16].

With the Maxwell construction (the equal-area rule), the M-line can be produced by the pressure equilibrium condition, $P^*(v_M) = P^*(v_L) = P^*(v_G) = P^{*S}$, and it can be expressed with a linear function in the temperature range considered: $\rho_M = 0.3073T^* - 0.09778$.

For applying Eq.(2), Eq.(8) and Eq.(9), the influence parameter, $m$, is required, which is a weak function of temperature or density. For the LJ fluid we adopt the values reported in Ref.[13] and the data are fitted with a linear function: $m = -0.6556T^* + 6.047$. This correlation is used to compute the influence parameter in the temperature range, $0.65 \le T^* \le 1.1$. Some calculation results are presented by Figure 1 – Figure 4, where the position variable is denoted as $x = z - z_0$ for brevity.

Figure 1a depicts definitions of the classic (the algebraic mean) dividing interface and the mean-field interface, Eq.(5). The mean-field dividing interface correctly reflects the fact that the decay length of the vapor side is less than that of the liquid side. In comparison, the algebraic mean interface wrongfully equal-divides the interface area.

Figure 1b presents the 1st and 2nd derivatives obtained from the mean-field density profile, Eq.(6) and the classic density profile, Eq.(4) at $T^* = 0.65$. This is a very important comparison. Apparently, the positional differences are caused by the different definitions for the dividing interface. Because Eq.(6) is physically favorable and more accurate than Eq.(4) (Figure 1c and Figure 2), the derivatives obtained from the classic model are not acceptable. Therefore predictions of the intrinsic properties from the classic model, Eq.(4), are not reliable.

Figure 1c illustrates correlation results for the density profiles over the entire temperature range, $0.65 \le T^* \le 1.1$, by the classic model, Eq.(4) (dashed lines), and by the new model, Eq.(6) (solid lines), respectively. The observations are: (1) the new model, Eq.(6), works excellently over the entire range from vapor phase to liquid phase; (2) the classic model, Eq.(4), is less accurate, in particularly on the deep liquid side. More detailed comparisons are provided in Ref.[16].

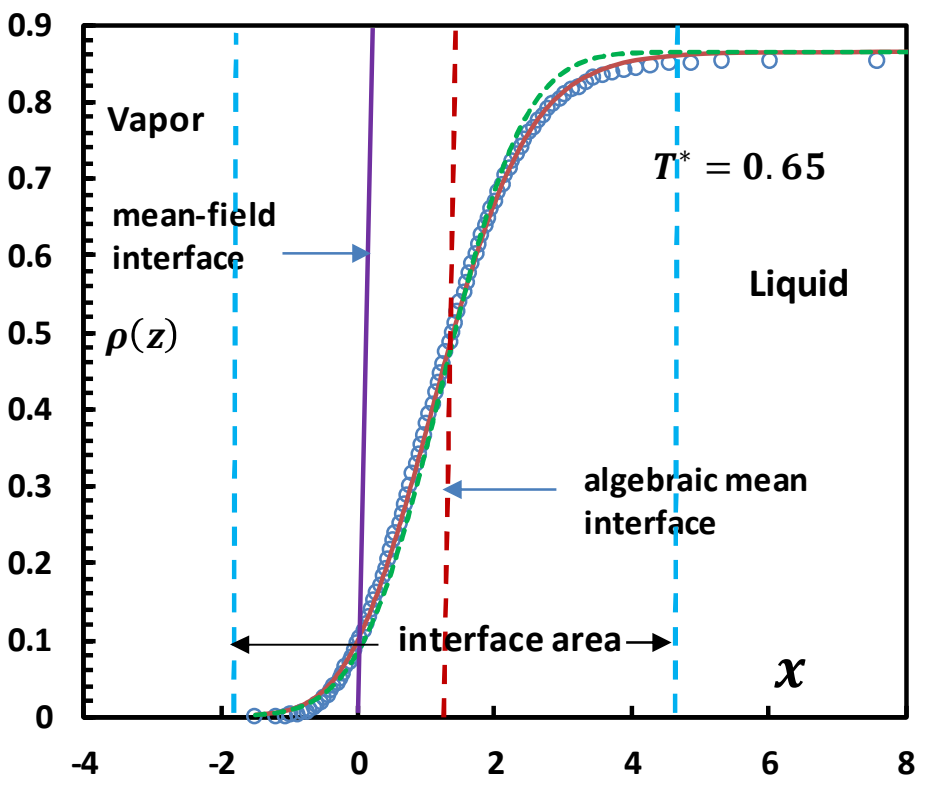

**Figure 1a**. Density profiles at $T^* = 0.65$, where $x = z - z_0$. The red solid curve is from Eq.(6), dashed green line from Eq.(4) and circles are from the most recent simulation data [17,20].

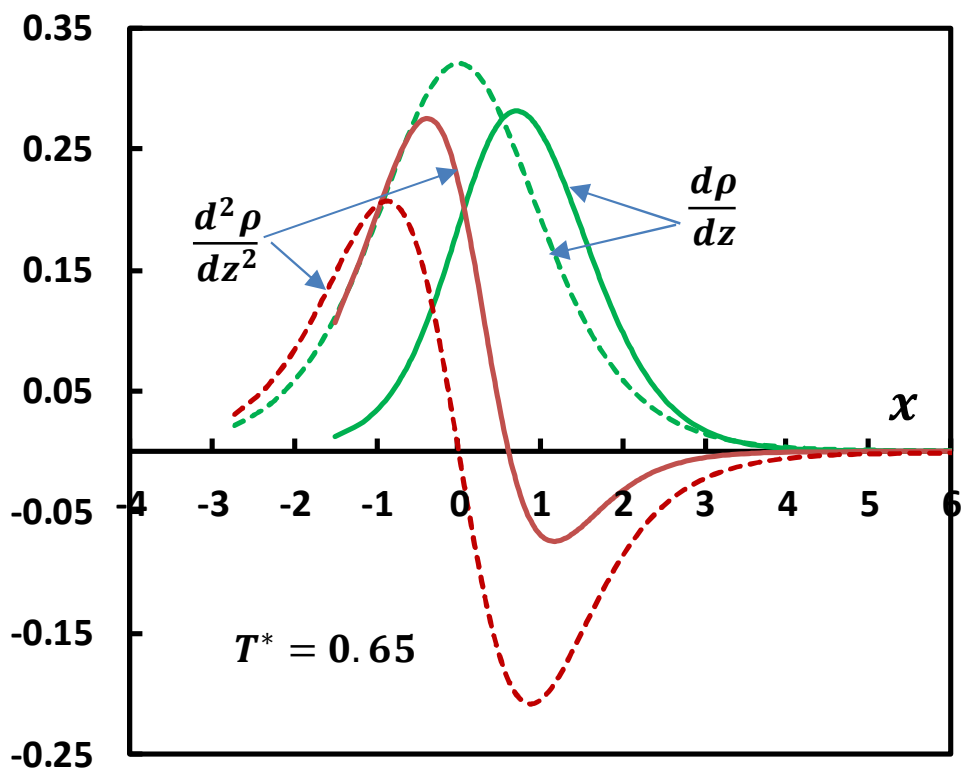

**Figure 1b**. Derivatives from two density profile models: solid lines are from Eq.(6) and dashed lines are from Eq.(4).

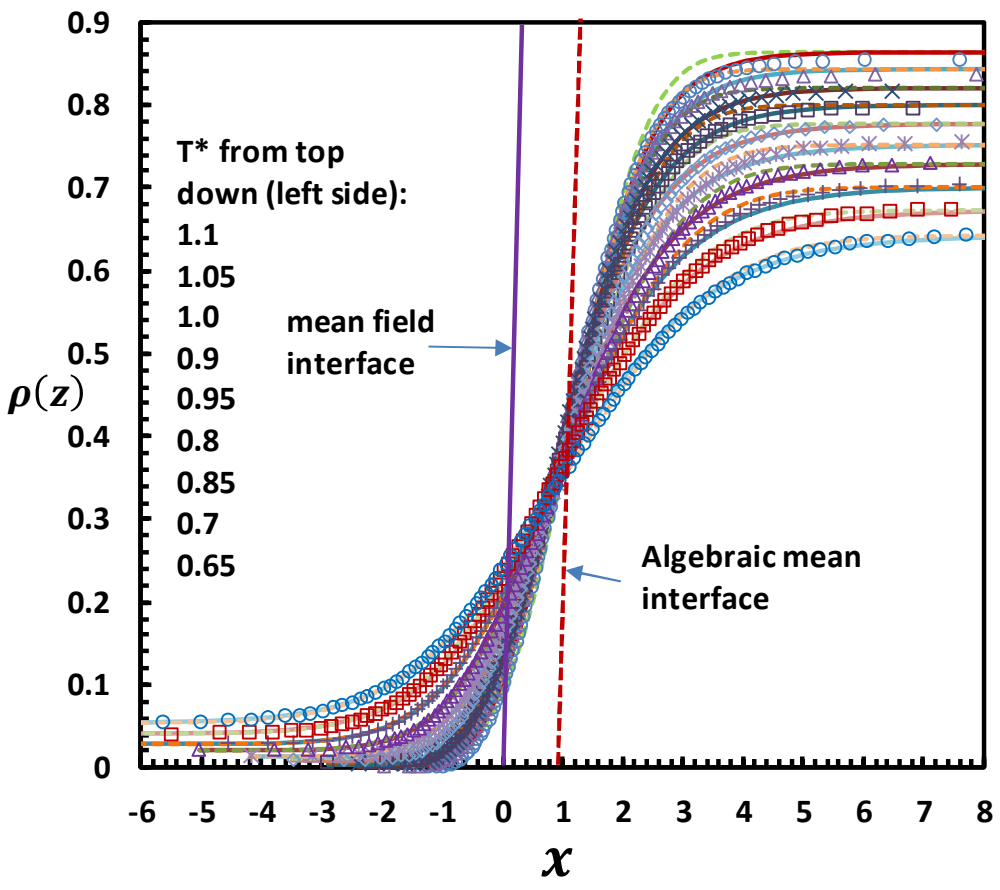

**Figure 1c**. Plots of density profiles in the entire range.

**Figure 1**. Definitions of the interfacial region and density profile.

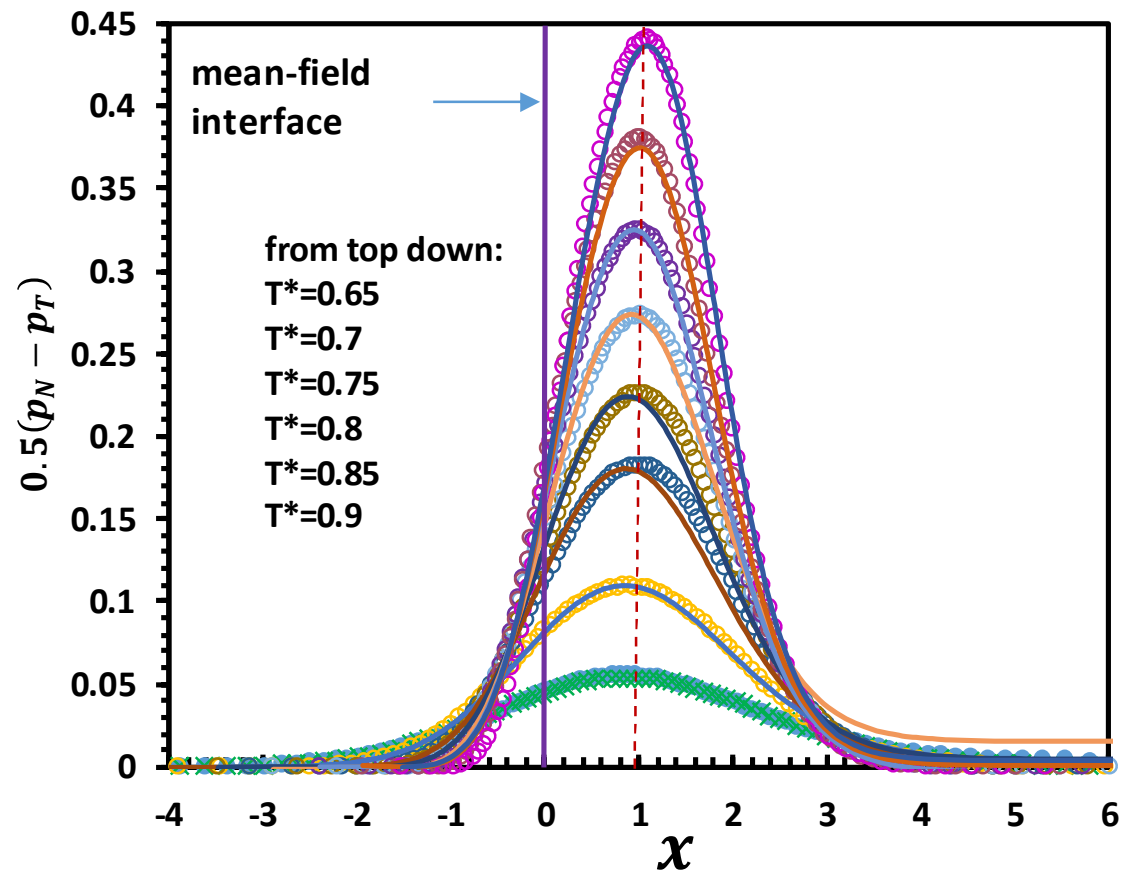

**Figure 2**. Plots of predictions of the pressure difference, $[p_N(z) - p_T(z)]/2$ by Eq.(3) (lines), compared with the simulation data [20]. The density profile is given by Eq.(6).

Figure 2 depicts the predictions for the pressure difference, $p_N(z) - p_T(z)$ given by Eq.(3) and Eq.(6). Both the values of the pressure difference and it's peaks are accurately predicted. In summary, the new density profile model has the following merits: (1) physically meaningful boundaries at the vapor and liquid sides; (2) excellent correlations of the density profile data; (3) highly accurate predictions of the pressure differences. All the above guarantee that the results obtained from the density gradient theory based on Eq.(6) are physically reliable.

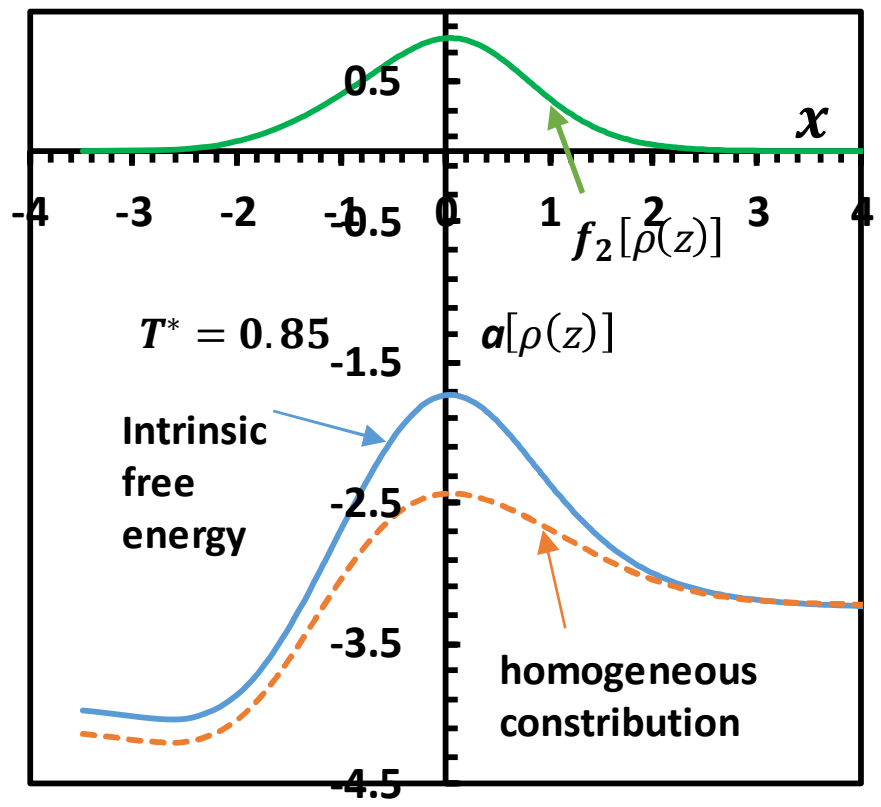

**Figure 3**. Helmholtz free energies at T*=0.85. The intrinsic free energy (solid blue line) is calculated by Eq.(8). Homogeneous contribution is from the LJ EoS [11] with the density $\rho(z)$. The green line is the heterogeneous contribution, $f_2(\rho(z))$.

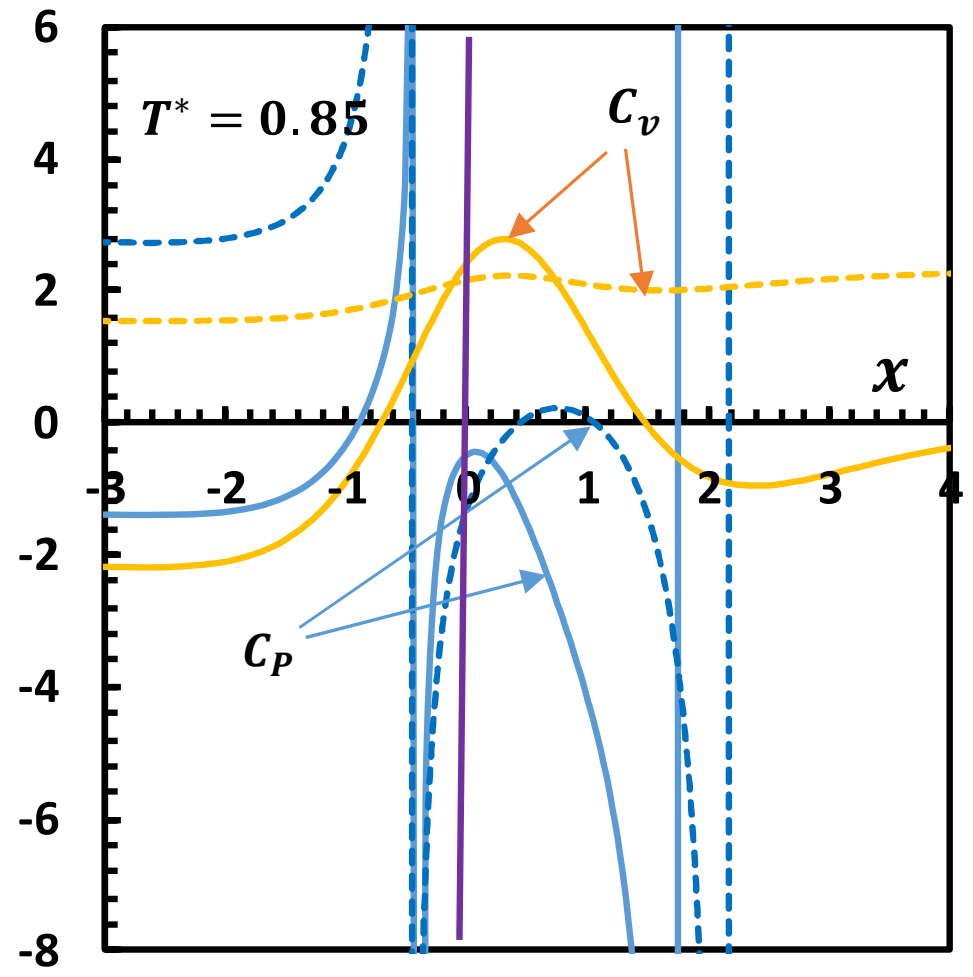

**Figure 4a** Heat capacities at T*=0.85. The solid lines depict the intrinsic heat capacities and dashed line, the homogeneous contributions.

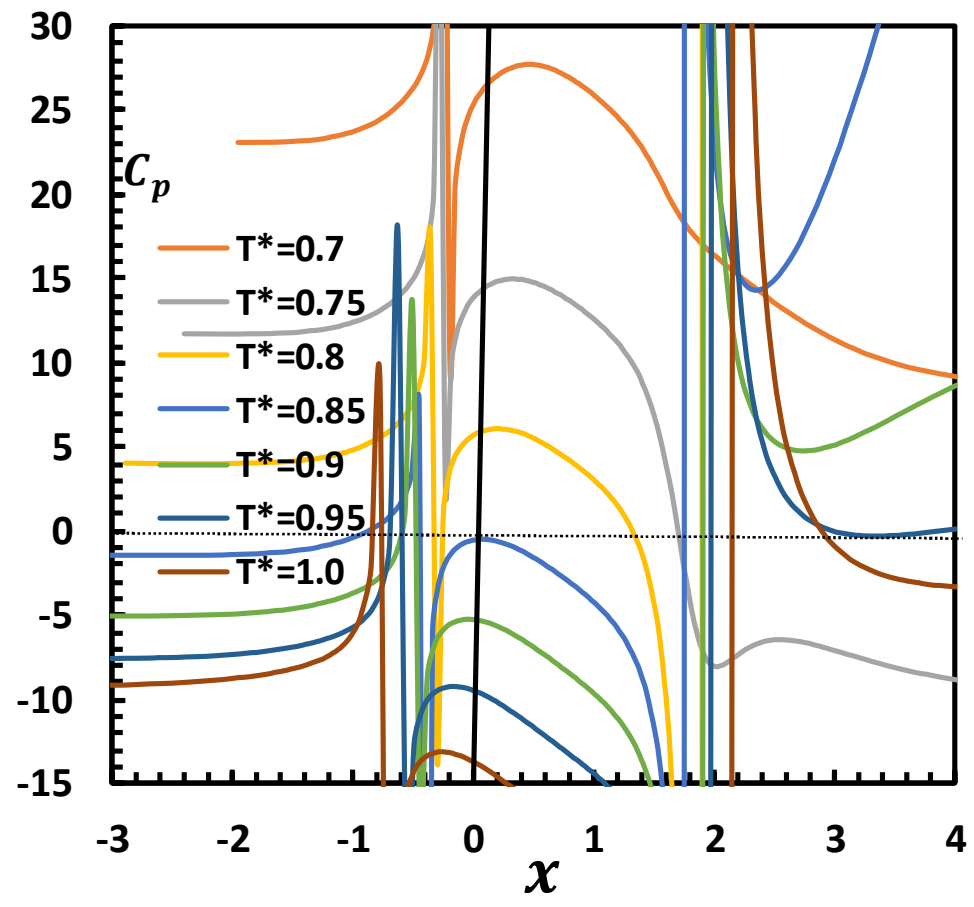

**Figure 4b**. Intrinsic isobaric heat capacity at multiple temperatures.

**Figure 4**. Plots of the local heat capacities of the LJ fluid.

Figure 3 illustrates the intrinsic free energy, Eq.(8), and it's components. An important observation is that the free energy peaks at the mean-field interface. This feature is responded with the local maximum of isobaric heat capacity, as shown by Figure 4a and Figure 4b.

Figure 4a plots the isobaric and isochoric heat capacities at T*=0.85. There is a transition temperature, $T_r \approx 0.65$: as $T_r < 0.65$, $C_P > 0$; as $0.65 < T_r < 1.0$, $C_P < 0$. In the stable fluid region, $C_P > C_v > 0$ while in the interfacial region, $C_P < C_v$. This implies that in the interfacial region temperature has a profound impact on the internal energy.

Figure 4b illustrates the intrinsic isobaric heat capacity at different temperatures. The most important observation is that the intrinsic heat capacity $C_P$ peaks (local maximum) at the mean-field interface. The slight deviations of the maxima from the interface are mostly caused by the discrepancies between the saturated densities from Ref.[17,20] (for determining $\rho(z)$ and $\tilde{f}_2$) and from Ref.(11] (the LJ EoS for calculating $C_v$ and $C_P$) [16]. The maximum locations of the intrinsic free energy and heat capacity do not coincide with the dividing interface when the classic model, Eq.(4), is employed [16].

Another observation is that at low temperature region ($T^* \leq 0.85$), $C_P > 0$. This finding is unexpected since previous theories or observations argue that in heterogeneous nanoscale systems the heat capacity is negative [21-23]. As temperature rises ($T^* > 0.85$), negative heat capacity is indeed observed.

Now we turn to the supercritical region. The focus is on the area where the Widom line divides it into the gas-like and liquid-like regions [8,9]. The Widom line is defined as the loci of the local maximum of the isobaric heat capacity [9]: $\left(\partial C_p/\partial T\right)_P = 0$. By using the LJ equation of state [11] and the heat capacity expression Eq.(9) in the supercritical region (where $\tilde{f}_2 = 0$ and density is uniform), the Widom line is generated [9,16]. With the Widom line, we can complete the phase diagrams in all the pressure-temperature-volume (density) planes.

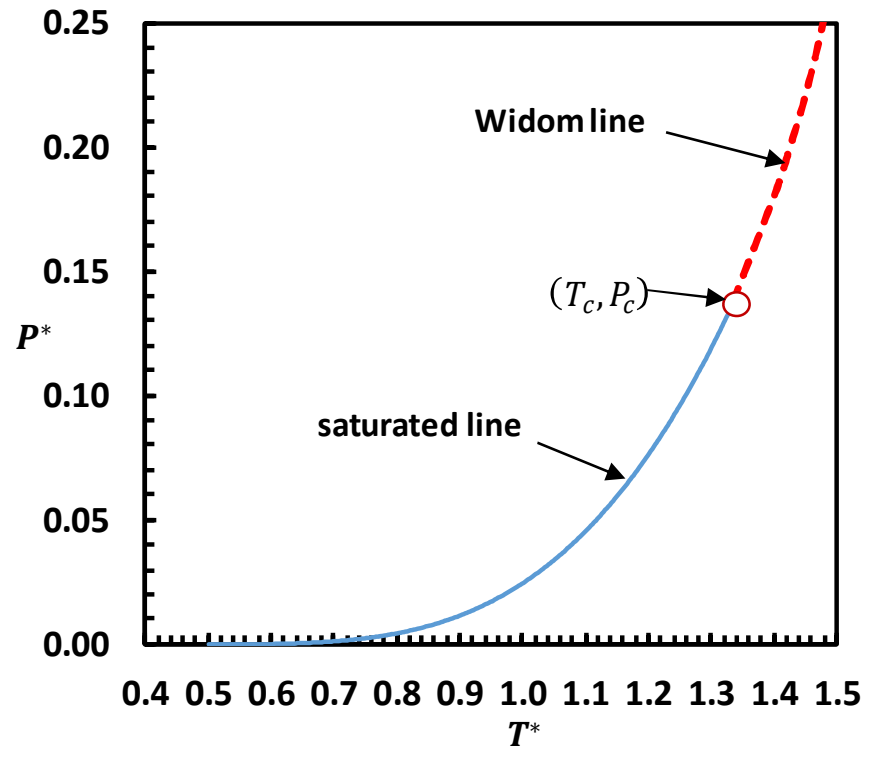

**Figure 5a**. Phase diagram in the temperature-pressure $(T^*, P^*)$ plane. The saturated curve is produced with the data from Ref.[11].

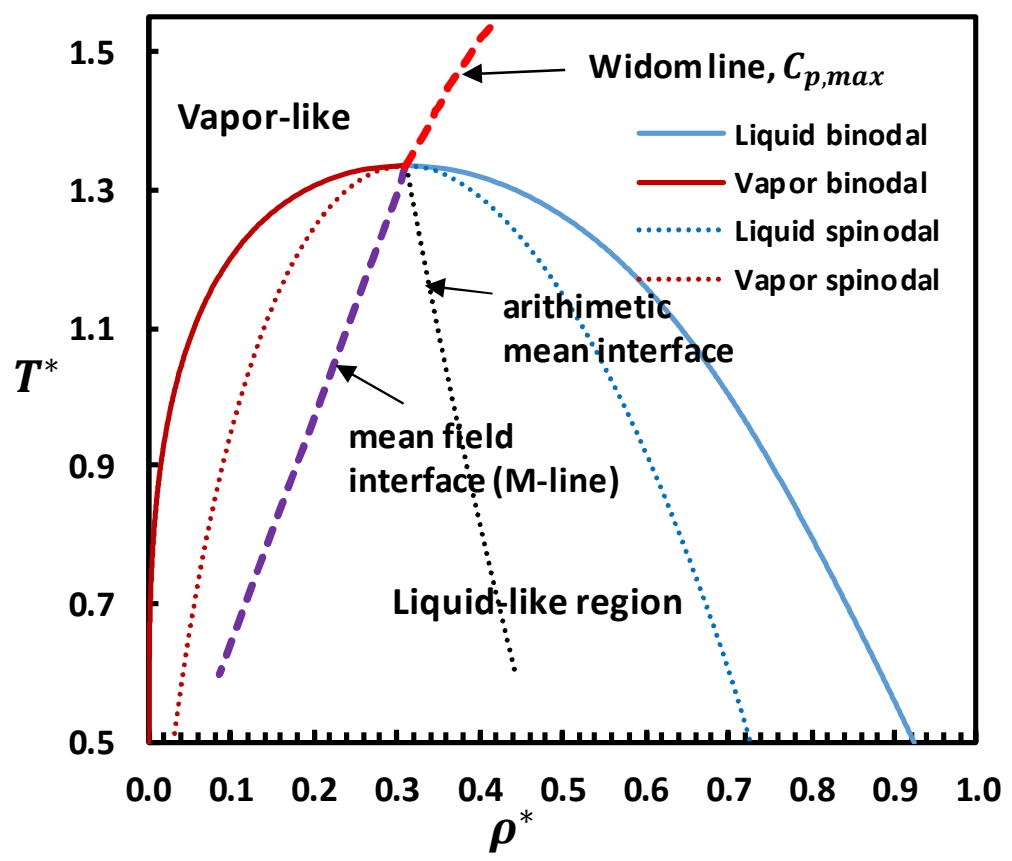

**Figure 5b**. Phase diagram in the temperature-density $(T^*, \rho^*)$ plane. The binodal and spinodal curves are produced with the data from Ref.[11].

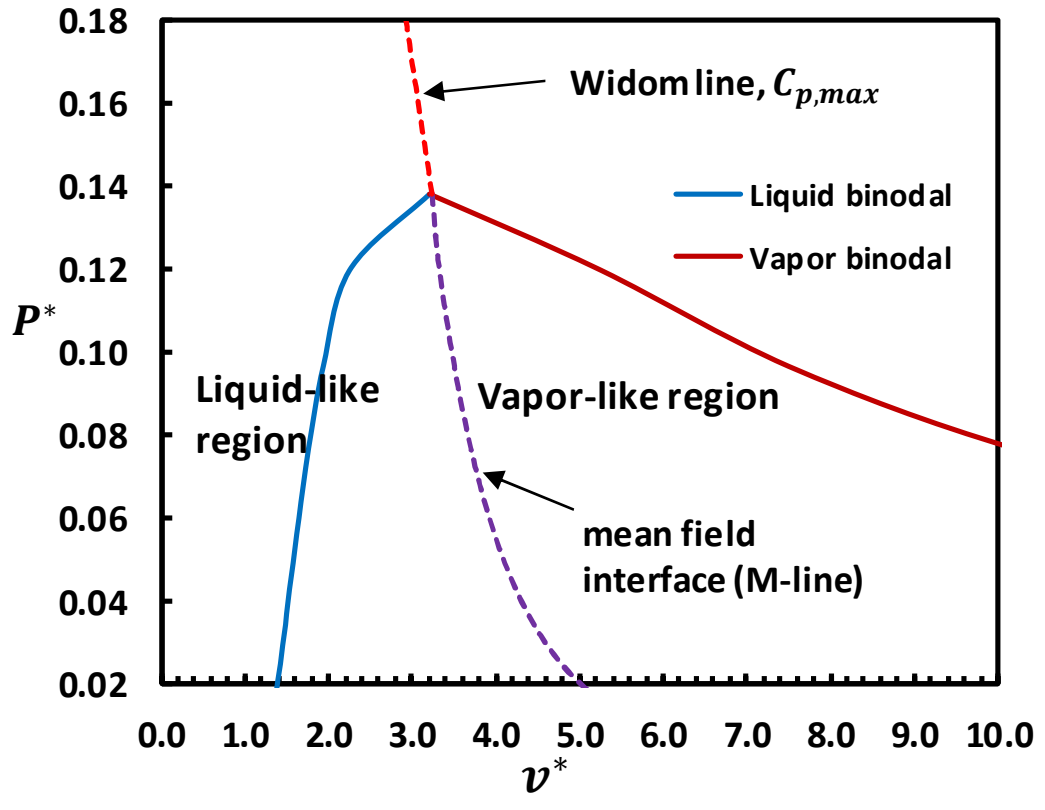

**Figure 5c**. Phase diagram in the volume-pressure $(v^*, P^*)$ plane.

**Figure 5.** Phase diagrams in all the planes

Figure 5 depicts the phase diagrams in all three planes. Figure 5b and Figure 5c show that the mean-field interface divides the interfacial area into vapor-like and liquid-like regions. This leads to our primary conclusion that the M-line is the natural extension of the Widom line into the coexistence region, or vice versa. Therefore the gas-like and liquid-like behaviors in the supercritical region are inherited from the coexistence region and the "single" line composed of the dividing interface and the Widom line functions as the unique demarcation line for the entire phase space. Here the "entire phase space" refers to the vapor-liquid coexistence region and the supercritical region where the Widom line functions as the dividing curve.

Another interesting feature is the sign change of the heat capacity. In a stable fluid, heat capacity at constant volume is related to the mean-square fluctuation of internal energy, $C_v = \overline{\Delta E^2}/k_B T^2$, and is thus always positive, and $C_P > C_v$. In the vapor-liquid coexistence region, as the system is composed of heterogeneous nanoscale clusters, the heat capacities exhibit negative values [21-23]. Some argue [22] that "negative heat capacity in nanoclusters is an artifact of applying equilibrium thermodynamic formalism to a cluster…". On the other hand, theoretical analysis [21] and experiment [23] reported negative heat capacities in some nanocluster systems. The results obtained from the density gradient theory and the mean-field theory for the LJ fluid add richer behaviors. In the interfacial region, $C_v > C_P$, and there is a "transition" temperature, $T^* \approx 0.85$ ($T_r \approx 0.65$), above which $C_P < 0$ and below which $C_P > 0$ (Figure 4b). Nevertheless, the feature of the local maximum of the isobaric heat capacity is reserved in the entire phase space.

This work also reveals that a mean-field EoS contains all the information for composing the phase diagrams over the entire phase space. All three solutions to the VLE problem (for a vast majority of EoS) have their roles or significances: two as the saturated volumes of vapor and liquid phases, respectively, which are related to the first-order transition and the third one as the dividing interface, related to the second-order transition. Finally, along the M-line, the critical point can be seen as the transition point where "static" nanoclusters become dynamic, indicated by the sign change of the intrinsic heat capacity from negative to positive. Here the term "static" does not exclude the size changes of the nanoclusters.

## Acknowledgements

The author is grateful to Dr. Langenbach and Dr. Stephan for helpful discussions and providing their research articles. Special thanks to Dr. Mejía for helpful discussions and kindly providing unpublished data of the density profile and the pressure differences, $p_N(z) - p_T(z)$ for the LJ fluid in the temperature range $0.65 \le T^* \le 1.1$.